# Competitive performance analysis of two evolutionary algorithms for routing optimization in graded network

Kavitha Sooda
Advanced Networking Research Group (RIIC), DSI
and Asst. Professor, Dept. of CSE
Nitte Meenakshi Institute of Technology
Bangalore - 560064, India
e-mail: kavithasooda@gmail.com

T. R. Gopalakrishnan Nair
ARAMCO Endowed Chair- Technology, PMU, KSA
Advanced Networking Research Group
VP, Research, Research and Industry Incubation Center
(RIIC), Dayananda Sagar Institutions,
Bangalore - 560078, India
e-mail: trgnair@ieee.org, www.trgnair.org

*Abstract*— **In this paper we compare the two intelligent route generation system and its performance capability in graded networks using Artificial Bee Colony (ABC) algorithm and Genetic Algorithm (GA). Both ABC and GA have found its importance in optimization technique for determining optimal path while routing operations in the network. The paper shows how ABC approach has been utilized for determining the optimal path based on bandwidth availability of the links and determines better quality paths over GA. Here the nodes participating in the routing are evaluated for their QoS metric. The nodes which satisfy the minimum threshold value of the metric are chosen and enabled to participate in routing. A quadrant is synthesized on the source as the centre and depending on which quadrant the destination node belongs to, a search for optimal path is performed. The simulation results show that ABC speeds up local minimum search convergence by around 60% as compared to GA with respect to traffic intensity, and opens the possibility for cognitive routing in future intelligent networks.**

## I. INTRODUCTION

In recent years Evolutionary algorithms has been applied for understanding the collective behavior of known population. Owing to the rapid increase in the end-users and new service demands in Internet, it has confronted with problems like scalability and performance issues along with increasing complexity as the network expands. There exist a requirement of Internet or future global network test-beds which support reliable end-to-end connectivity for federation and management of Internet. On this complex structure of Internet, there still exists a challenge of routing and imbibing intelligence among the routers. Current routing strategy whether it is wired or wireless routing protocols, all of them are based on some predetermined calculation or prior knowledge of the connection between the nodes. This information was simply based on the request packet sent. Generally for any intelligent action [1] to take place a collective behavior of the topology and elements, need to be studied prior to the path determination process. Also there exists a need for enabling a self organizing feature of the topology when changes occur. Such scenario where there is a requirement for decision-taking process by router, evolutionary intelligence [2-3] has played a pre-dominant role in the recent past. It is a self- adaptive technique which can be applied for routing where the desired result is obtained based on the exchange of the information by the agents.

The remainder of the paper is organized as follows: Section 2 outlines the overview of Network Awareness and Graded network. Section 3 deals with ABC algorithm and GA. Section 4 present the details of the mathematical model. Section 5 consists of simulation results. Finally, the last section concludes the paper.

## II. NETWORK AWARENESS AND GRADED NETWORK

### A. Network Awareness

The awareness of a network helps in taking decision for routing, as it creates an understanding of the nodal-link setup and its pairing performance better. It supplements the knowledge of network characteristics when deciding on the forward channels to route. There exists a need for intelligent querying which is required by the nodes to participate in network routing. Existing intelligent routing technique works well for small scale systems. Thus creation of awareness of the environment will make the routing easy for largely distributed nodes as discussed in [4]. The exploitation of the environmental knowledge is aided by the topological setup and bandwidth availability. The nodes are queried for identification of better ones which is based on the observed facts. Here the observed facts include the network QoS metrics like node density (ND), resource allocation (RA), traffic congestion (TC), network lifetime (NL), delay and bandwidth. Querying here is to test whether a given QoS metric for the node is above or below a threshold value. This helps in identifying the nodes which are feasible for routing. The algorithm, initially assume that any given node is a feasible candidate for obtaining the optimal path. Decision is taken on nodes based on the query operator which will help to bifurcate the non-feasible nodes. Next the neighbor of the previous feasible node is considered and the process is repeated until the destination node is reached. Here the network aware algorithm is based on exploration of the neighbor nodes as illustrated in a simple way in Fig. 1.

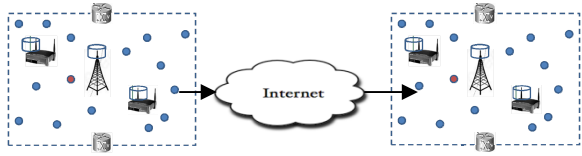

Fig. 1. Network Awareness Routing

Thus the environment awareness [5] can extract the nodes which are feasible to propagate the data in an efficient manner. This leads to an improved performance due to reduction in the search space and provides feasible path for the data to reach the desired destination in an efficient manner.

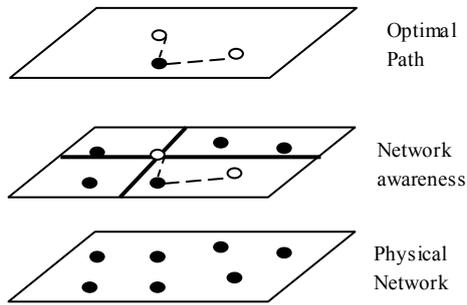

Fig. 2. Network Awareness algorithm

The network awareness algorithm is applied on a physical network where the nodes which participate in routing are available. Now a quadrant is designed where the quad is abstracted in vitrual circuit plane with source as the centre point as shown in Fig. 2. Here the nodes which belong to the destination quadrant will participate in routing. Thus the optimal path is determined assuming sufficient nodes are available in selected quadrant of interest.

### B. Graded network

The concept of graded network has been discussed by the authors in [6-8]. Grade value estimation method and grade surface creation for implementing intelligent routing in the autonomic network is a core focus of research. Grade is like a coagulated index reflection for intelligent systems to work in network; it is made available everywhere and routing will much depend on it. It signifies the quality of the router, which is in fact the knowledge of the environment. The router must be an intelligent entity because it performs different operations based on the environment information. It depends on input, output, load and resource availability. The graded network is setup with six QoS parameters. There are considered at two levels; level-1 and level-2. *Level-1* is applied region-wise and the goal is to achieve favorable routing based on selected attributes. The values obtained from *Level-1* must be able to eliminate the non-production node. Non-production nodes are those that come in the pitfall of congestion and possess less resource availability. The grade approach must be able to exact the QoS satisfied nodes from the network setup. The selected nodes are assigned values that indicate how efficient they are. At this point, we are able to calibrate the nodes for the routing process region-wise. The algorithm used to determine the optimal path requires proactive decision making on the output obtained. This is because many paths exist to reach the destination, and we have to choose the most optimal path. Once the gradient value has been calculated, it can be made available as pervasive information packets to all the other nodes to obtain the optimal path constitution.

For the first level of node selection we apply the *level-1* operation. The nodes are prioritized based on local observation. Prioritizing is performed based on ontological reasoning as shown in Fig. 3:

```
IF (NL) {
    IF (ND < 5) {
        IF (TC does not exist){
            IF (RA){
                IF(Delay does
                not exist){
                    P=1;
                ELSE
                    P=2;
            ELSE
                P=3;
        ELSE
            P=4
    ELSE
        P=5;
ELSE
P=6;
```
Fig. 3. Priority Model

Based on the priority values, nodes which have value greater than or equal to three are selected for routing.

*Delay calculation*

We assume that the node location, external traffic requirements $\gamma_{ik}$, channel cost $d_i(C_i)$, the constants D, μ and the flow ($\lambda_i$) are given and feasible. Thus the Delay T at the node, is given by (1).

The average rate of message flow $\lambda_i$, on the $i^{th}$ channel is equal to sum of the average message flow rate of all paths that traverse this channel. The traffic entering the network

$$T = \sum_{i=1}^{M} \frac{\lambda i}{\gamma} \left[ \frac{1}{\mu C i - \lambda i} \right]$$

(1)

from the external sources forms a Poisson's process with a mean $\gamma_{jk}$ (messages per second) for those messages originating at node j and destined for node k. Therefore total external traffic entering and leaving the network are considered to be equal to $\lambda_i$.

Now the graded function, i.e., *Level-2*, considers the calculated values as its input of *level-1*. The output of this function defines the route availability for the set of nodes considered. This shall be calculated for all the available paths leading towards the destination node. The mean value of the gradient in the grade function shows the success level of operation of the network, which is based on the bandwidth. Calculation for traffic congestion and bandwidth availability are further explored in section 4.

## III. BEE COLONY ALGORITHMS AND GENETIC ALGORITHM

In recent years swarm intelligence has been the focus of interest in research. Swarm intelligence is about developing an algorithm or a distributed problem solving device inspired by the collective behavior of social insect colonies and other animals.

### A. Artificial Bee Colony(ABC) for network routing

Artificial Bee colony algorithm is developed by Karaboga in 2005 [10], which was motivated by the intelligent behavior of the bees. It is an optimization tool, where search is performed in a distributed environment. Here the artificial bee fly and find the food source based on experience or their nest mates and adjusts their position. Some artificial bees (scouts) choose the food source randomly without experience. If the nectar amount of the new food source is better, then the bees memorize this position and forget the previous food source position. Thus ABC performs a local search with onlooker and employed bees, with global search methods managed by the onlooker and scout bees, attempting to balance exploration and exploitation process.

The following shows the main steps in the algorithm:

Send scouts to the initial food source.
REPEAT
    Step 1: Employed bees are sent to the food sources to determine the nectar amount.
    Step 2: Based on the preference of the onlooker bees calculate the probability value of the sources.
    Step 3: Send onlooker bees to the food source and determine the nectar amount.
    Step 4: The sources which are exhausted with the nectar, are not exploited further.
    Step 5: Randomly send the scouts for discovering new food sources.
    Step 6: Remember the best food source obtained so far.
UNTIL (requirement is met)

### B. Genetic algorithms

Genetic algorithms are a part of evolutionary computing which has been used as a powerful optimization technique for finding near optimum solution is large search space. It is also an efficient search method that has been used for path selection in networks [12].
The algorithm is as follows:
**begin** PATHSELECTION_GA
    Create initial population of n nodes randomly.
    **while** generation_count < k do
        /* k = max. no. of generations.*/
    **begin**
        Selection
        Fitness Function
        Modified crossover
        Mutation
        Increment generation_count.
    **end** ;
Output the optimal path by selecting the highest probability value chromosome on which data can be sent.
**end** PATHSELECTION_GA.

ABC outperformance GA for the following three reasons:

i) GA produces new solutions based on the present solution at hand whereas in ABC produces candidate solution from taking the difference between randomly determined part of the parent and a randomly chosen solution from its population. This helps in the convergence speed of search into a local minimum.

ii) Diversity in solution is obtained by mutation operation in GA whereas in ABC a new one is randomly inserted by a scout. This helps is global search and prevents premature convergence.

iii) In GA there are three control parameters, i.e, crossover rate, mutation rate and number of generation where is ABC there is only one control parameter, i.e, limit with fixed colony size and maximum cycle number.

### C. Fitness function

The simulation topology is generated randomly. Here relaxation is imposed on the number of onlooker bees. The number of onlooker bees is equal to the number of neighbors to the source node who belong to the quadrant of the destination node. Here the fitness value is calculated on nodes which belong to the quadrant. The nodes which satisfy the threshold bandwidth values are the ones which get selected and participate in routing which indicates the fitness value. Further the authors have planned to implement multi-objective parameters solvable by the fitness function.

## IV. MATHEMATICAL MODEL

The Fig. 4 shows how the network evolves itself to determine the optimal path. Here we have proposed a channel level stochastic Markov process for realizing the link capacity for random network model. Every region is considered to be a *Jackson network* [13] in this model. Here the set of n links, L={1,2,3, …,n} are assumed to have the same capacity, where l ∈ L.

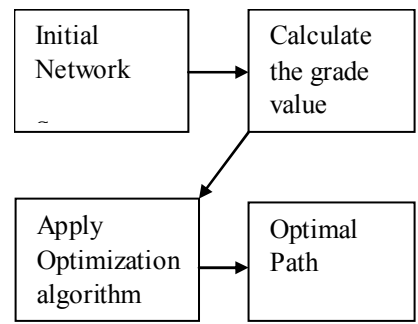

Fig. 4. Network Setup

Let the flow arrival rate between any two nodes, say s and r be represented by $\lambda_{sr}$. Along the route where packet flows we assume that each flow transmits packet at fixed unit rate with mean $\mu^{-1}$.
The traffic intensity of the link can be calculated as,

Traffic Intensity = $\frac{P_s \cdot T_l(t)}{A_b}$ (4)

Here Ps indicates the packet size, Ab indicates the available bandwidth and the traffic load on the link l ∈ L at time t is denoted by $T_l(t)$. This represents the total number of flows currently routed across it. Link l's flow arrival rate is denoted by $\gamma_l$. This flow represents a part of the flow arrival between *s* and *e* that is routed to link l. Here the flow departure rate is proportional to the current link load $T_l(t)$ and is given by $T_l(t) * \mu$. The bandwidth availability on the link and packet size for the simulation is fixed.

The updates on the link is done every 30*t sec, which would represent the delay involved in building a grade network. Here the model is built based on the given information on traffic load and flow rate. Here we assume that the traffic arrivals during [0, t] are routed based on the link states $T_l(0)$ available at time 0. The link state $T_l(t)$ tracks to the following differential equations which derives the total traffic arrival rate $\gamma_l$ to link l.

$\frac{d\,T_l(t)}{dt} = (\gamma_l - T_l(t) * \mu)$ (5)

This is of the form,
$$\frac{dy}{dx} + Ry = S$$
Therefore the Integrating Factor(IF) is,
Y(IF) = ∫S(IF)dt +c
Hence eq 5 becomes,

$T_l(t) = T_l(0)\,e^{-\mu t} + \gamma_l / \mu [1 - e^{-\mu t}]$ (6)

where $T_l(0)$ is the link state information at time interval [0,t]. This value is subtracted from link capacity to obtain the remaining fraction of the bandwidth available.

$B_a = L_c - T_l(0)$; (7)

This output is considered by the level-2 operation for calculating the available bandwidth. The traffic intensity is thus determined from Eq.(4), using Eq. (6) and Eq. (7). Equation 7, helps in the calculation of whether congestion has occurred or not. Here we have TC determined for the priority model.

The model is designed based on nodes and link generated at random. To each node, the jobs arrive from outside which are Poisson in process with rate α >0. Each arrival is independently routed to node j with probability $p_j \geq 0$ and $\sum_{j=1}^{k} P_{ij} = 1$. Here k represents the total number of packets which has arrived from node i.

Thus the determination of the link capacity helps to identify where congestion has occurred or not. The agent here is used for aiding the grade network with relevant information when required. For instance at the first level of grading the parameters which are looked for are network lifetime, congestion level check, delay, node density and resource allocation. All these five parameters are determined by the agent. For instance the network lifetime is randomly assigned by the system for simulation purpose. The agent finds the remaining lifetime of the node and check whether it is above a threshold value. The bandwidth availability is determined by Eq. (7) which in turn helps to check whether congestion has occurred or not. Delay is obtained by Eq. (4). Network node density is number of packets arriving to the node and resource allocation is randomly assigned.

We then apply local algorithm trying to achieve global optimization, which is referred to as constrained optimization. Based on agent information, we applying the probabilistic reasoning approach to reduce congestion. Here the agent tries to do the following:

Minimize $\sum_{j \in N(i)}^{n} |(\text{Act } T_l(j) - \text{Cur } T_l(j))|$
Subject to $\sum_{j \in N(i)}^{n} \text{Act } T_l(j) \geq E$
Over $0 \leq T_l(j) \leq 1$ for all j

Here Act $T_l(j)$ represents the actual bandwidth, Cur $T_l(j)$ represents the current bandwidth. The summation of actual bandwidth must be ≥ envisaged bandwidth for a particular route in the direction of destination which is considered by the level-2 operation.

Upon this selected nodes the optimization algorithm are applied to compare the performance of GA and ABC for determining the optimal path.

## V. SIMULATION RESULTS

The initial configuration of simulation model consists of a randomly distributed set of nodes and links in a two dimensional search space as shown in Fig. 5. Now propagating message, demands user to enter the source and destination node for which a quadrant appears with source as the centre. Depending on the destination node quad, the grade values of the nodes are determined. As mentioned in Eq. 1 and Eq.7, the delay and bandwidth availability will be calculated and made available in the knowledge base. The quadrant structure reduces the search space nearly to one-fourth depending on the source location. Now the optimal path determination is performed only in the quadrant where destination belongs to. The paths are obtained with ABC and GA algorithms with initial parameters as shown in table 1.

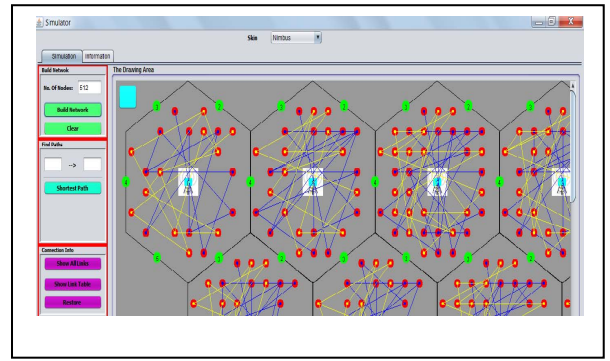

Fig. 5. Simulation Snapshot

The grade value shown remains same for execution of both the algorithms. The design of the nodes and link setup takes care that there is no much randomness exist in the search space.

Table. II shows the hop count derived from the path obtained after applying ABC and GA on the quadrant where destination belongs to. The values obtained here is one of the random run carried out.

TABLE I.   INITIAL PARAMETER

| Initial Parameter | Value |
|---|---|
| Population Size/Colony size | 15-1024 |
| Number of Iteration/ Maximum Cycle Number | 30 |
| Selection, crossover and mutation in GA | Roulette wheel, two-point crossover with 0.1% mutation |
| Packet size | 200 Bytes |
| Maximum Bandwidth | 30 Mbps |

The column two of the table shows the reduction in the number of nodes participating in the optimal path determination with grade value.

Here the number of onlooker bees is chosen based on the number of neighbors the source possess.

TABLE II.   HOP COUNT

| Total nodes | Number of nodes selected | Graded ABC Route length | Graded GA Route length |
|---|---|---|---|
| 15 | 10 | 3 | 4 |
| 16 | 12 | 4 | 3 |
| 32 | 18 | 5 | 8 |
| 64 | 38 | 5 | 7 |
| 128 | 61 | 6 | 8 |
| 256 | 64 | Path not available | Path not available |
| 512 | 234 | 7 | 7 |
| 1024 | 432 | 6 | 7 |

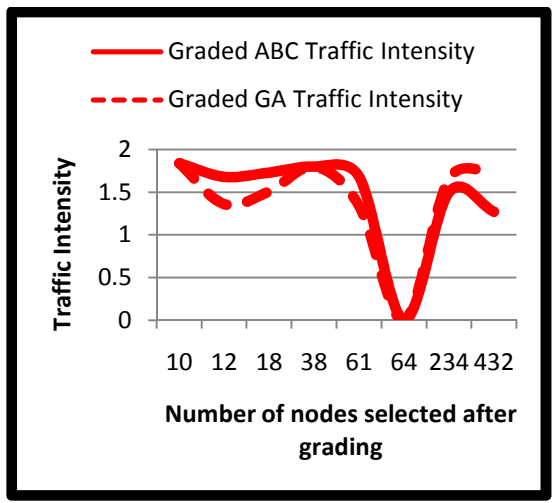

Fig. 6. Graph determining the Traffic Intensity in ABC and GA.

The employed bees now search for the best nectar among the available neighbors for each one of them. This search is continued until the destination is reached. Now among the different paths which exist between the source and destination one complete path is selected based on the bandwidth estimation of the links. Since here the links selected have bandwidth above a threshold value, there is less likely scenario for congestion to occur.

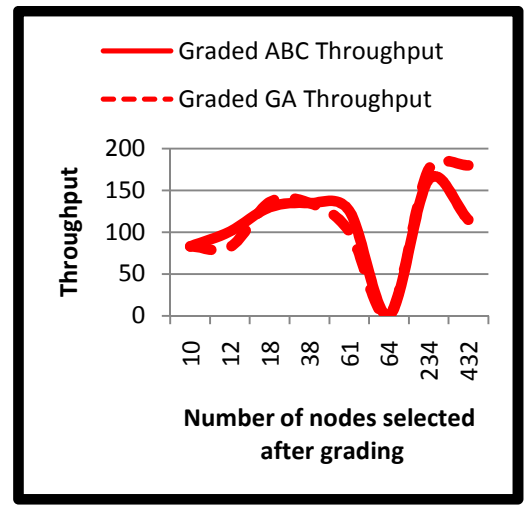

Fig. 7. Graph determining the throughput in ABC and GA.

From Fig. 6, we observe that the convergence speed for ABC is 60% better than GA. The graph in Fig. 7, shows that most of the path obtained by ABC was of better grade value hence the link selected were of higher bandwidth. Since the work is based on the quadrant design we may not always get the best paths by ABC or GA. The technique applied here assures reduction in the search space and helps in convergence of path faster with quality. From Table. II, we saw that they may not exist fruitful path in the quadrant. We also observed that GA learnt better quality paths 10% of the time as compared to ABC and 10% equal quality paths.

## VI. CONCLUSION AND FUTURE WORK

The paper here presented the optimal path determination using the evolutionary artificial bee colony and genetic algorithm approaches in general network. ABC showed a good performance in terms of path convergence rate as the control parameter of the algorithm was only total number of cycle as against to GA which had three control parameters. A quality path with sufficient bandwidth was determined over the graded network. Here we have shown that without a pre-defined message path we are able to obtain the paths much optimally realizing a multi-hop connecting route. The routers are able to dynamically select the forward nodes and propagate the messages successfully to the destination. However the simulation has certain limitations as it is on progressive path. Currently only few hundreds of nodes were used to test and evaluate the principles of the model.

Future work involves expanding the simulation of higher number of nodes realizing near real life network scenarios. The simulation setup shall be modified to a zone based structure where there shall be a zone coordinator which will be responsible improving scalability and for achieving parallelism. The remaining two self organization properties shall be further implemented where parallel search techniques would be explored for intelligent state space update.

REFERENCES

[1] T. R. Gopalakrishnan Nair, Kavitha Sooda, Abhijith, "Transformation of networks using cognitive approaches", JRI, Vol.1, No. 1, December, 2008, pp.7-14.

[2] H.F. Wedde, M. Farooq, Y. Zhang, Beehive: an efficient fault-tolerant routing algorithm inspired by honeybee behavior, ant colony, optimization and swarm intelligence, in: 4th International Workshop, ANTS 2004, Brussels, Belgium, September 5–8.

[3] J. Kennedy, R.C. Eberhart, "Particle Swarm Optimization", IEEE International Conference on Neural Networks, vol. 4, 1995, pp. 1942–1948.

[4] T R Gopalakrishnan Nair, M Jayalalitha, S Abhijith, "Cognitive Routing with Stretched Network Awareness through Hidden Markov Model Learning at Router Level", IEEE Workshop on Machine learning in Cognitive networks, Hong Kong, 2008.

[5] Yanif Ahmad Ugur Cetintemel "Network-Aware Query Processing for Stream-based Applications", Proceedings of the 30th VLDB Conference, Toronto, Canada, 2004.

[6] T. R. Gopalakrishnan Nair, Kavitha Sooda, "Particle Swarm Optimization for Realizing Intelligent Routing in Networks with Quality Grading", WiCOM, 7th IEEE International Conference on Wireless Communications, Networking and Computing, Wuhan, China, 23-25th September, pp. 1-4, 2011.

[7] T. R. Gopalakrishnan Nair, Kavitha Sooda," Application of Genetic Algorithm on Quality Graded Networks for Intelligent Routing", World Congress on Information and Communication Technologies-2011, December 2011, Mumbai, India, pp. 558-563, DOI: 10.1109/WICT.2011.6141306.

[8] Kavitha Sooda, T. R. Gopalakrishnan Nair,"A Comparative Analysis for Determining the Optimal Path Using PSO and GA", International Journal of Computer Applications, Volume 32– No.4, October 2011, pp no. 8-12.

[9] V. Tereshko, Reaction–diffusion model of a honeybee colony's foraging behaviour, in: M. Schoenauer (Ed.), Parallel Problem Solving from Nature VI, Lecture Notes in Computer Science, vol. 1917, Springer–Verlag, Berlin, 2000, pp. 807–816.

[10] D. Karaboga, An idea based on honeybee swarm for numerical optimization, Technical Report TR06, Erciyes University, Engineering Faculty, Computer Engineering Department, 2005.

[11] D. Karaboga, B. Basturk, On the performance of artificial bee colony (ABC) algorithm, Applied Soft Computing 8 (1) (2008), pp.687–697.

[12] Shengxiang Yang, Hui Cheng and Fang Wang, " Genetic Algorithms with Immigrants and memory schemes for dynamic shortest path routing problems in mobile AdHoc networks", IEEE Transactions on systems, man, and cybernetics – applications and reviews, vol. 40, no. 1, Jan 2010.

[13] J. Medhi, *Stochastic Models in Queueing theory*, Elsevier, Academic press, 2006.